\documentclass[final,5p,times,twocolumn]{elsarticle}

\usepackage[T1]{fontenc}

\usepackage{amssymb}
\usepackage{amsmath,mathtools}
\usepackage{subcaption}
\usepackage{bm}
\usepackage{xspace}

\usepackage{lineno}

\newcommand{\TypeINA}{Type-I $\bar{N}A$\xspace}
\newcommand{\TypeIIpA}{Type-II $\bar{p}A$\xspace}
\newcommand{\TypeIInA}{Type-II $\bar{n}A$\xspace}
\newcommand{\Ca}[1]{${}^{#1}\mathrm{Ca}$\xspace}

\journal{Physics Letters B}
\biboptions{sort&compress}

\begin{document}

\begin{frontmatter}



\title{Toward testing antinucleon--nucleus optical potentials with antineutron scattering lengths}


\author[first]{Hiroyuki Fujioka\corref{cor1}}
\cortext[cor1]{Corresponding author.}
\ead{fujioka@phys.sci.isct.ac.jp}
\author[first]{Sayaka Ishii}
\affiliation[first]{organization={Department of Physics, Institute of Science Tokyo},
            addressline={2-12-1 Ookayama}, 
            city={Meguro},
            postcode={152-8551}, 
            state={Tokyo},
            country={Japan}}

\begin{abstract}
Low-energy antineutron--nucleus scattering would provide experimental constraints on the complex $s$-wave scattering length.
Using charge symmetry, we calculate scattering lengths from optical potentials constrained by antiprotonic-atom data and examine their dependence on the potential strength, finite interaction range, and nucleon density distributions.
An analytic approximation for strongly absorptive two-parameter Fermi potentials clarifies the different roles of the potential geometry and complex strength in determining the real and imaginary parts.
Numerical calculations show that the neutron surface profile affects the predicted scattering lengths, particularly the imaginary part.
As a case study, we examine an isovector extension motivated by a recent theoretical analysis of antiprotonic calcium data.
The isovector term increases $\operatorname{Re}a$ by $0.3\text{--}0.4\,\mathrm{fm}$ and $-\operatorname{Im}a$ by $0.4\text{--}0.5\,\mathrm{fm}$ for \Ca{48}, while the spread among different nucleon density distributions is magnified by the isovector term.
These results provide quantitative predictions for comparison with future low-energy elastic and annihilation measurements.
Such measurements would independently constrain the threshold interaction and test whether a common optical-potential parametrization can describe antineutron scattering across several nuclides.
\end{abstract}




\end{frontmatter}




\section{Introduction}

Low-energy hadronic interactions have been a central subject of hadron physics for decades.
Hadron--hadron interactions have been studied through both scattering measurements and femtoscopic correlation measurements.
Although the hadron--nucleus interaction is built up from the elementary hadron--nucleon interaction,
additional complexity arises from in-medium effects and nuclear structure.

In general, the two-body short-range interaction without the Coulomb interaction is encoded in the scattering amplitude.
In the low-energy limit, only the $s$-wave ($\ell=0$) contributes to the scattering amplitude, and the $s$-wave scattering amplitude $f_0$ is related to the $s$-wave phase shift $\delta_0$ by
\begin{align}
  f_0=\frac{\exp(2i\delta_0)-1}{2ik}=\frac{1}{k\cot\delta_0-ik},
\end{align} 
where $k$ is the center-of-mass wave number.
The effective range expansion,
\begin{align}
  k\cot \delta_0=-\frac{1}{a}+\frac{1}{2}r_ek^2+\cdots,\label{eq:effective_range_expansion}
\end{align}
defines the scattering length $a$ and the effective range $r_e$.
For example, neutron scattering below epithermal energies is characterized by the scattering length alone, and the neutron absorption cross section is given by $4\pi (-\operatorname{Im}a)/k$, known as the $1/v$ law~\cite{Sears1992}.

Unlike neutron scattering, the direct determination of the scattering length for other hadronic probes is difficult because of the lack of sufficiently low-energy hadron beams and the dominant Coulomb interaction for charged hadrons.
Instead, the shift and width of the 1s state of exotic atoms, in which an electron is replaced by a negatively charged hadron, yield the scattering length through the Deser--Trueman relation~\cite{Deser1954-sq,Trueman1961-aq}.
However, this method is in practice limited to exotic hydrogen and deuterium, since strong absorption prevents the atomic cascade in heavier systems from reaching the \(1s\) state.

For antineutrons, one of the neutral hadrons, a scheme for producing low-energy beams with momenta as low as
$9\,\mathrm{MeV}/c$ has recently been proposed~\cite{Filippi2025,Amsler:2930906}.
The antineutron can be produced via the charge-exchange reaction ($\bar{p}p\to \bar{n}n$) by backward scattering in the center-of-mass frame using $300\,\mathrm{MeV}/c$ antiproton beams that might be supplied by the Antiproton Decelerator (AD) at CERN.
Antineutrons of unprecedentedly low energy will offer unique opportunities to explore antineutron physics~\cite{Filippi2025,Filippi2026-hs}, including measurements of antineutron scattering off nuclei in the $s$-wave regime.
The elastic and annihilation cross sections of antineutrons with nuclei, denoted as $\sigma_\mathrm{el}$ and $\sigma_\mathrm{ann}$, respectively, are related to the complex scattering length $a=a_\mathrm{R}-ia_\mathrm{I}$ as a function of the center-of-mass wave number $k$:
\begin{align}
  \sigma_\mathrm{el} &= 4\pi |a|^2(1-2a_\mathrm{I}k)+\mathcal{O}(k^2), \label{eq:sigma_el}\\
  \sigma_\mathrm{ann} &= \frac{4\pi}{k}a_\mathrm{I}-8\pi a_\mathrm{I}^2+\mathcal{O}(k). \label{eq:sigma_ann}
\end{align}
The contributions of higher partial waves ($\ell>0$) are expected to be suppressed because a characteristic interaction radius $R$ (approximately the nuclear radius) satisfies $kR<1/2$. 
Thus, the low-energy antineutron--nucleus scattering experiment is expected to provide a direct determination of the antineutron--nucleus scattering length\footnote{According to Eqs.~(\ref{eq:sigma_el}) and (\ref{eq:sigma_ann}), the cross sections determine $a_I$ and $|a_R|$, but not the sign of $a_R$.
However, the strongly absorptive potentials considered here yield a positive $a_R$.}.
For comparison, the minimum momentum for which antineutron--nucleus annihilation cross-section data are available~\cite{Astrua2002-ez} is $76\,\mathrm{MeV}/c$, too high to evaluate the $s$-wave scattering length.

At present, even the antinucleon--nucleon scattering lengths are poorly constrained, owing to the admixture of partial waves in $\bar{p}p$ scattering and to unresolved hyperfine splitting in protonium~\cite{Carbonell2023-lt,Gotta2004-bq}.
Meanwhile, the \textit{ab initio} description of $\bar{p}+d$, $\bar{p}+{}^3\mathrm{H}/{}^3\mathrm{He}$, and $\bar{p}+{}^4\mathrm{He}$ systems has advanced rapidly~\cite{Lazauskas2021-wp,Duerinck2023-qw,Duerinck2026-mv,Dehghani2026-sk,Dehghani2026-pq}, and measured antineutron--light-nucleus scattering lengths would provide a valuable benchmark for such calculations.

For heavier nuclei, the antinucleon--nucleus interaction has been investigated for decades in the optical-potential model~\cite{Batty1997-nd,Friedman2007-vu},
which has been applied to pion--nucleus and antikaon--nucleus systems as well.
The optical-potential parameters were phenomenologically fitted to level shifts and widths of antiprotonic atoms.
The direct determination of the scattering lengths for various nuclides would also serve as a critical test of the optical-potential model.

In this Letter, we calculate antineutron--nucleus scattering lengths from optical potentials constrained by antiprotonic-atom data and examine their dependence on the potential strength, finite interaction range, and nucleon density distributions.
An isovector extension motivated by the isotope dependence of antiprotonic calcium spectra is considered to illustrate the sensitivity of the predictions to the optical-potential parametrization.
The key question is whether such optical potentials can account for scattering lengths extracted from future low-energy antineutron measurements.
Measurements across several nuclides would further test whether a common set of optical-potential parameters can describe the near-threshold interaction.

The remainder of this Letter is organized as follows.
Section~\ref{sec:optical_potential} introduces the antiproton--nucleus optical potential and the corresponding antineutron--nucleus potential, via charge symmetry.
In Sec.~\ref{sec:antinucleon_scattering_length}, we calculate the scattering length for a two-parameter Fermi potential and compare it with an analytic approximation; the scattering length as a function of the mass number is then deduced by assuming nucleon density distributions.
We extend this analysis to realistic density distributions obtained from density functional theory in Sec.~\ref{sec:isovector_potential}, examining the role of the isovector term for ${}^{40}\mathrm{Ca}$ and ${}^{48}\mathrm{Ca}$.
Finally, Sec.~\ref{sec:summary} summarizes our findings and discusses the prospects for a direct measurement of the antineutron--nucleus scattering length.

\section{Antiproton--nucleus optical potential}\label{sec:optical_potential}

Antiprotonic atoms, in which an electron is replaced by a negatively charged antiproton, have been used to investigate antiproton--nucleus interactions.
Strong-interaction shifts and widths of atomic levels, deduced from X-ray spectroscopy, constrain the optical potential between an antiproton and a nucleus.
The following $s$-wave optical potential in the low-density approximation is commonly used~\cite{Batty1997-nd,Friedman2007-vu}:
\begin{align}
  2\mu V_\mathrm{opt}^{(\bar{p}A)}(r)=-4\pi\left(1+\frac{\mu}{m}\frac{A-1}{A}\right)[b_0\rho(r)+b_1\delta\rho(r)],\label{eq:optical_potential_full}
\end{align}
where $\mu$ is the reduced mass, $m$ is the nucleon mass, $A$ is the mass number of the nucleus, and $\rho(r)=\rho_n(r)+\rho_p(r)$ ($\delta \rho(r)=\rho_n(r)-\rho_p(r)$) is the isoscalar (isovector) density distribution.
The coefficients $b_0$ and $b_1$ are complex parameters corresponding to in-medium antiproton--nucleon isoscalar and isovector scattering lengths\footnote{We note that $b_0$, $b_1$ and the scattering length $a$ defined in Eq.~(\ref{eq:effective_range_expansion}) each follow the sign convention adopted in the respective literature, which results in an opposite overall sign between them. The overall minus sign on the r.h.s. of Eq.~(\ref{eq:optical_potential_full}) compensates for the difference in sign conventions.}, respectively.

Global fits to antiprotonic-atom data have shown that the isovector and $p$-wave coefficients are consistent with zero~\cite{Batty1995-cb,Friedman2005-yl}.
The isoscalar form obtained by setting $b_1=0$ in Eq.~(\ref{eq:optical_potential_full}),
\begin{align}
  2\mu V_\mathrm{opt}^{(\bar{p}A)}(r)=-4\pi\left(1+\frac{\mu}{m}\frac{A-1}{A}\right)b_0\rho(r),\label{eq:optical_potential_b0only}
\end{align}
has been used to extract the neutron density distribution in the surface region~\cite{Friedman2005-yl}.
However, a recent calculation of atomic levels for antiprotonic ${}^{40,48}\mathrm{Ca}$ atoms revealed that either an isovector or a $p$-wave term is necessary to reproduce the isotope dependence of the shifts and widths~\cite{Yoshimura2025-vt}. We will incorporate the $s$-wave isovector term in Sec.~\ref{sec:isovector_potential}.

Antineutron--nucleus ($\bar{n}A$) and antiproton--nucleus ($\bar{p}A$) optical potentials are related to each other by charge symmetry, which is the invariance under a rotation by $\pi$ about the $I_y$ axis in isospin space, interchanging $p\leftrightarrow n$ and $\bar p\leftrightarrow\bar n$.
Applying charge symmetry to Eq.~(\ref{eq:optical_potential_full}) gives  the antineutron--nucleus optical potential: 
\begin{align}
  2\mu V_\mathrm{opt}^{(\bar{n}A)}(r)=-4\pi\left(1+\frac{\mu}{m}\frac{A-1}{A}\right)[b_0\rho(r)-b_1\delta\rho(r)].\label{eq:optical_potential_antineutron}
\end{align}
In particular, for a nuclide with $N>Z$, the isovector term enters the antineutron--nucleus and antiproton--nucleus interactions with opposite signs.
Hence, the assumption of $b_1=0$ made so far should be carefully reexamined.

\subsection{Examples of optical potentials}
Batty~et al.~\cite{Batty1995-cb} determined the optical potential parameters by fitting to the experimental data from various antiprotonic atoms. Using the form of Eq.~(\ref{eq:optical_potential_full}), the following parameters were obtained:
\begin{align}
  b_0=(2.51\pm 0.24)+(3.46\pm 0.27)i\,\mathrm{fm},\quad b_1=0,\label{eq:optical_potential_unfolded}
\end{align}
which we refer to as ``\mbox{Batty1995}''.
They also examined extended forms containing an isovector term, a $p$-wave term, and a density-dependent term; here we adopt the simplest zero-range, isoscalar version so as to compare it directly with the parametrization of Ref.~\cite{Friedman2005-yl} below.

The global fit was updated in 2005~\cite{Friedman2005-yl} by including antiprotonic-atom X-ray and radiochemical data from the PS209 experiment at CERN~\cite{Trzcinska2001-mp}.
Whereas a zero-range interaction was assumed in the previous global fit~\cite{Batty1995-cb}, Friedman et al.~introduced a folded density distribution $\rho^F(r)$ to account for the finite range of the antiproton--nucleon interaction:
\begin{align}
  \rho^F(r)=\int d\bm{r'}\rho(\bm{r'})\frac{1}{\pi^{3/2}\beta^3}\exp\left[-\frac{(\bm{r}-\bm{r'})^2}{\beta^2}\right],\label{eq:folding}
\end{align}
with the range parameter $\beta=0.85\,\mathrm{fm}$.
This yielded the parameters:
\begin{align}
  b_0=(1.3\pm 0.1)+(1.9\pm 0.1)i\,\mathrm{fm},\quad b_1=0.\label{eq:optical_potential_folded}
\end{align}
Simultaneously, they determined the parameters of the neutron density distribution, which minimized the $\chi^2$ of the global fit. This optical potential with finite-range folding is referred to as ``\mbox{Friedman2005}''.

Equation~(\ref{eq:optical_potential_b0only}) with the above $b_0$ parameter serves as the standard optical potential for low-energy antinucleon physics.
For example, it underlies the realistic evaluation of the intranuclear suppression factor for neutron--antineutron oscillation in ${}^{16}\mathrm{O}$ nuclei~\cite{Friedman2008-sf}, which converts the antineutron-appearance lifetime limits of Super-Kamiokande into a bound on the free neutron--antineutron oscillation time $\tau_{n\bar{n}}$~\cite{Abe2021-td}. 

In related work, the same type of optical potential has also been applied to studies of antiproton annihilation in the nuclear periphery, a process sensitive to the peripheral neutron-to-proton density ratio.
The PS209 collaboration used this approach to systematically investigate the neutron density distributions of stable nuclei, in particular the neutron skin thickness~\cite{Trzcinska2001-fn} (see Sec.~\ref{sec:mass_no_dependence}).
This approach will be extended to short-lived nuclei by the PUMA experiment~\cite{Aumann2022-qk}.

\subsection{Applicability to very light nuclei}
The application of global optical-potential parametrizations to very light nuclei is not straightforward, as illustrated by the difficulties encountered in describing antiprotonic helium and low-energy $\bar{p}\text{--}{}^{4}\mathrm{He}$ annihilation within a common parametrization~\cite{Batty2001-hh}.
We therefore restrict the present calculations with the optical potentials of Batty1995 and Friedman2005 to nuclei with $A\ge 12$. 

\section{Antinucleon--nucleus scattering length}\label{sec:antinucleon_scattering_length}

In Ref.~\cite{Batty1983-sn}, the $s$-wave scattering lengths were evaluated from the optical potential proposed for exotic atoms, such as antiprotonic atoms.
The antiproton--nucleus strong-interaction scattering length was calculated by solving the Schr\"odinger equation with the optical potential, but without the Coulomb potential, at a very low energy.
In the following, the sign of the scattering length of Ref.~\cite{Batty1983-sn} has been inverted to match the convention of this Letter.
The author deduced the scattering lengths for 13 nuclides, using the shift and width of the corresponding antiprotonic atoms, and found that the real part of the scattering length is approximately proportional to $A^{1/3}$, whereas the imaginary part is almost constant:
\begin{align}
  a=(1.52\pm 0.01)A^{0.313\pm 0.017}-(1.11\pm 0.06)i\,\mathrm{fm}.
\end{align}
This result was compared with that for a square-well potential: in the limit of strong absorption, the real part would be equal to the potential radius, and the imaginary part would vanish.
The finite imaginary part was attributed to the finite diffuseness of the potential.
The result was updated by incorporating more data on antiprotonic atomic levels~\cite{Batty2001-hh}, and the following strong-interaction scattering length was obtained: 
\begin{align}
  a=(1.54\pm 0.03)A^{0.311\pm 0.005}-(1.00\pm 0.04)i\,\mathrm{fm}.\label{scattering_length_Batty2001}
\end{align}
However, a further systematic scattering-length analysis based on the latest measurement by the PS209 collaboration~\cite{Trzcinska2001-mp} has not been reported.
Instead, antineutron--nucleus scattering lengths were evaluated from the optical potential in Ref.~\cite{Friedman2005-yl} for $^{12}\mathrm{C}$ and $^{58}\mathrm{Ni}$~\cite{Friedman2008-sf}.

Under charge symmetry and the additional assumption of $b_1=0$, the above strong-interaction scattering length can be directly identified with the antineutron--nucleus scattering length.
This antineutron--nucleus scattering length serves as the only input currently available for antineutron interactions with bulk material in terms of Fermi's pseudopotential.
This quantity is critically important in discussions of the experimental feasibility of neutron--antineutron oscillation searches~\cite{Nesvizhevsky2019-vw,Gudkov2020-dn,Protasov2020-gy,Shima2025-ie,Fujioka2026-qp}.

In the following, the scattering length for a given optical potential is calculated numerically using the variable phase approach~\cite{Calogero1967}.

\subsection{Analytic approximation to the scattering length for a 2pF potential}\label{sec:analytic_approximation}

\begin{figure*}[t!]
  \centering
  \includegraphics[width=\textwidth]{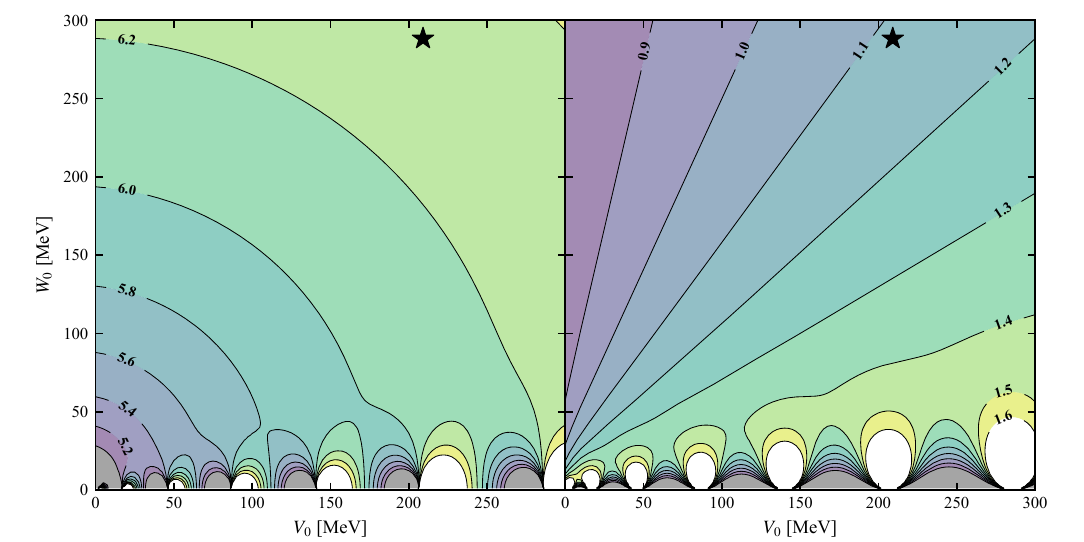}
  \caption{Scattering length for 2pF potentials, $\operatorname{Re}a$ (left) and $-\operatorname{Im}a$ (right) in units of fm for a potential depth of $-(V_0+iW_0)$, with the half-density radius and the diffuseness fixed to $5\,\mathrm{fm}$ and $0.5\,\mathrm{fm}$, respectively. The star represents the depth of the Batty1995 potential assuming the normal nuclear density $\rho_0=0.16\,\mathrm{fm}^{-3}$.}\label{fig:scattering_length_2pF}
\end{figure*}

To interpret the dependence of the scattering length on the potential strength and geometry, we consider the two-parameter Fermi (2pF) potential
\begin{align}
  V(r)=-\frac{U_0}{1+\exp[(r-c)/z]},
\end{align}
where $c$ is the half-density radius, $z$ is the diffuseness, and $-U_0$ is the complex potential depth.
We also define
\begin{align}
  K=\frac{\sqrt{2\mu U_0}}{\hbar},\qquad \zeta=-izK.
\end{align}
Figure~\ref{fig:scattering_length_2pF} shows the scattering length $a$ for 2pF potentials\footnote{The potentials considered in this subsection are not tied to a specific target nucleus. For these model-potential examples, we therefore approximate the reduced mass by the nucleon mass, $\mu=m$.} with the potential depth of $-(V_0+iW_0)$ and the geometric parameters of $c=5\,\mathrm{fm}$ and $z=0.5\,\mathrm{fm}$.
A similar plot for the imaginary part was already shown in Ref.~\cite{Protasov2020-gy}.
For this potential, the scattering length admits the following analytic approximation in the strongly absorptive regime:
\begin{align}
  a\approx c+2\gamma z+2z\ln \zeta-\frac{z}{6\zeta^2}\equiv a_\text{approx},\label{eq:approx_scattering_length}
\end{align}
where $\gamma=0.577\ldots$ is Euler's constant,
for $z\ll c$ and $(\operatorname{Im}K)c\gg 1$.
The derivation will be given in \ref{sec:scattering_length_2pF}.
For reference, the deviation of the scattering length $a_\text{approx}$ from the numerically calculated one is depicted in Fig.~\ref{fig:approx_deviation_2pF}, which shows
the approximation is valid to within 1\% accuracy for a sufficiently absorptive potential.
Under the simple assumption that the potential depth $U_0$ and the diffuseness $z$ are independent of the nuclide, $a_\text{approx}$ reduces to the half-density radius $c$ plus a nuclide-independent complex constant within the approximation of $\mu=m$.
Conversely, a deviation from this behavior might indicate the breakdown of the simple 2pF optical potential with the fixed depth and diffuseness.

To decompose $a_\text{approx}$ into its real and imaginary parts, we write $U_0=|U_0|\exp(i\varphi)$, where $|U_0|$ is the modulus and $\varphi$ is the argument ($0<\varphi<\pi/2$ for an attractive and absorptive potential, $\pi/2<\varphi<\pi$ for a repulsive and absorptive potential).
Since
\begin{align}
  \zeta&=-iz|K|\exp(i\varphi/2)=z|K|\exp(i\varphi/2-i\pi/2)
\end{align}
lies in the fourth quadrant of the complex plane, it follows that
\begin{align}
  \ln \zeta&=\ln(z|K|)+\frac{i}{2}(\varphi-\pi).\label{eq:logzeta}
\end{align}
The approximated scattering length $a_\text{approx}$ in Eq.~(\ref{eq:approx_scattering_length}) separates into real and imaginary parts:
\begin{align}
  \operatorname{Re} a_\text{approx} &= c+2\gamma z+2z\ln(z|K|)+\frac{\hbar^2}{12zm|U_0|}\cos\varphi\label{eq:approx_scattering_length_re}\\
  -\operatorname{Im} a_\text{approx} &= z(\pi-\varphi)+\frac{\hbar^2}{12zm|U_0|}\sin\varphi.\label{eq:approx_scattering_length_im}
\end{align}

The first two terms of $\operatorname{Re} a_\text{approx}$ depend only on the geometric parameters $c$ and $z$, while the potential depth $U_0$ enters through the remaining terms.
The almost concentric contours in the left panel of Fig.~\ref{fig:scattering_length_2pF} are due to the third term.
On the other hand, the first term of $-\operatorname{Im} a_\text{approx}$ depends on the argument of the potential depth, and it appears as almost radial contours in the right panel of Fig.~\ref{fig:scattering_length_2pF}.
This asymptotic form is already known for an exponential potential, $\propto \exp(-r/z)$~\cite{Karmanov2000-tm,Protasov2020-gy}.
We have shown that it is indeed a good approximation for a 2pF potential as well.
As $|\zeta|\approx [z/(0.5\,\mathrm{fm})]\sqrt{|U_0|/(82.9\,\mathrm{MeV})}$ is roughly unity, the last term in Eq.~(\ref{eq:approx_scattering_length}) is of order $10^{-2}\,\mathrm{fm}$; however, it cannot be neglected when 1\%-level accuracy is needed.

Although Eq.~(\ref{eq:approx_scattering_length}) was derived for the 2pF potential, it remains useful for a qualitative understanding of the scattering length for the folded potential (and a more complicated potential) as well.
Since the leading terms (except for the last one) in Eqs.~(\ref{eq:approx_scattering_length_re}) and~(\ref{eq:approx_scattering_length_im}) both increase with $z$, the larger surface diffuseness of a folded potential
is expected to increase both $\operatorname{Re}a$ and $-\operatorname{Im}a$
relative to the unfolded one.

\subsection{Mass-number dependence of the scattering length}\label{sec:mass_no_dependence}

\begin{figure}[t!]
  \centering
  \includegraphics[width=\columnwidth]{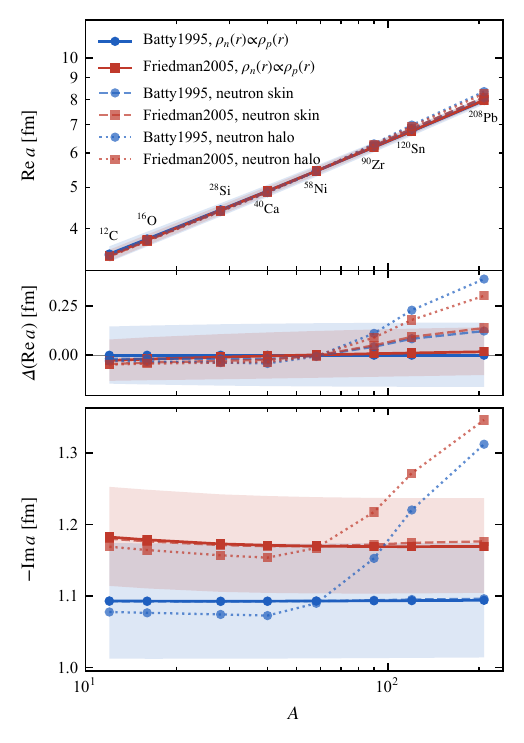}
  \caption{Scattering length for the \mbox{Batty1995} and \mbox{Friedman2005} potentials as a function of the mass number $A$, shown by blue and red lines, respectively. Solid, dashed, and dotted lines correspond to three different assumptions on the neutron density distribution, $\rho_n(r)\propto \rho_p(r)$ ($c_n=c_p$, $z_n=z_p$), neutron skin, and neutron halo in the terminology of Ref.~\cite{Trzcinska2001-fn}. $\varDelta(\operatorname{Re}a)$ is the deviation from the scattering length for the \mbox{Batty1995} potential with $\rho_n\propto \rho_p$. The diffuseness is fixed to be $z_p=0.49\,\mathrm{fm}$. The variation due to its standard deviation $0.037\,\mathrm{fm}$ is indicated by shaded bands.}\label{a_vs_A}
\end{figure}

In this subsection, we revisit the mass-number dependence of the scattering length evaluated using the optical potential.
For simplicity, we consider two types of potentials (Eq.~(\ref{eq:optical_potential_b0only})): ``\mbox{Batty1995}'' and ``\mbox{Friedman2005}'', whose parameters were given in Eq.~(\ref{eq:optical_potential_unfolded}) and Eq.~(\ref{eq:optical_potential_folded}), respectively.
These potentials do not contain the isovector term ($b_1=0$).
The parameters of the proton density distribution $\rho_p(r)$ are taken from Ref.~\cite{Wu2025-gi}. 
The authors fitted the half-density radius as a function of $A^{1/3}$: $c_p=1.232A^{1/3}-0.612\,\mathrm{fm}$. The diffuseness $z_p$ is set to $0.49\,\mathrm{fm}$.
Much less is known about the neutron density distribution $\rho_n(r)$.
The neutron skin thickness $\varDelta r_{np}=\sqrt{\langle r_n^2\rangle\vphantom{\langle r_p^2\rangle}}-\sqrt{\langle r_p^2\rangle}$ is assumed to obey $\varDelta r_{np}=-0.04+1.01(N-Z)/A\,\mathrm{fm}$~\cite{Trzcinska2001-fn}.
Two extreme cases, neutron skin ($c_n>c_p$, $z_n=z_p$) and neutron halo
($c_n=c_p$, $z_n>z_p$), are considered, with $z_n$ and $c_n$ chosen so as to reproduce the rms radius of the neutron density distribution.

Figure~\ref{a_vs_A} shows the scattering length as a function of $A$ for the \mbox{Batty1995} and \mbox{Friedman2005} potentials.
For each potential, three different assumptions on the neutron density distribution, $\rho_n(r)\propto \rho_p(r)$ ($c_n=c_p$, $z_n=z_p$), neutron skin, and neutron halo, are adopted.
The real part shows an $A^{1/3}$ dependence, as previous studies~\cite{Batty1983-sn,Batty2001-hh} already found.
More specifically, the $\varDelta(\operatorname{Re}a)$ panel reveals that, for each assumption on $\rho_n(r)$, the real part of the scattering length is almost the same for the two potentials\footnote{This does not mean that $\operatorname{Re}a$ is insensitive to $b_0$, since Batty1995 and Friedman2005 differ simultaneously in both $b_0$ and the interaction range $\beta$. These two changes move $\operatorname{Re}a$ in opposite directions from \mbox{Batty1995} to \mbox{Friedman2005}, and their effects partly cancel.}; rather, the nucleon density distribution is the dominant factor.

For heavy nuclides with $N>Z$, a finite $\varDelta r_{np}$ increases the real part of the scattering length; in particular, the real part is larger for neutron halo than for neutron skin.
This arises from the $z$-dependent term $(2\gamma+2\ln\zeta)z$ in the approximated expression Eq.~(\ref{eq:approx_scattering_length}).
The shaded bands in Fig.~\ref{a_vs_A} illustrate the sensitivity of the scattering length to $z_p$: it spans the range of values obtained by shifting $z_p$ by $\pm0.037\,\mathrm{fm}$, the standard deviation of $z_p$ across the nuclear chart reported in Ref.~\cite{Wu2025-gi}.
This band should not be read as the experimental uncertainty on $z_p$ for a specific nuclide; rather, it quantifies how strongly $a$ responds to nuclide-to-nuclide  variations in diffuseness.

In contrast, $-\operatorname{Im}a$ is larger for the \mbox{Friedman2005} potential than for the \mbox{Batty1995} potential, reflecting its larger effective diffuseness in the surface region, as already discussed above.
For the same reason, the variation of the diffuseness $z_p$ propagates almost linearly into $-\operatorname{Im}a$.
As the imaginary part is insensitive to the radial size of the potential, it is almost the same for neutron skin as for $\rho_n(r)\propto \rho_p(r)$, whereas neutron halo results in a substantially larger $-\operatorname{Im}a$.

Conversely, a systematic measurement of the scattering length across the periodic table up to ${}^{208}\mathrm{Pb}$ may shed light on the shape of the neutron density distribution, which is relevant to the density dependence of the symmetry energy~\cite{Vinas2014-aw}, provided that
$b_1$ is independently constrained.

The parametrization of Eq.~(\ref{scattering_length_Batty2001}), which is widely adopted in the literature~\cite{Nesvizhevsky2019-vw,Gudkov2020-dn,Protasov2020-gy,Shima2025-ie,Fujioka2026-qp}, is not directly comparable with the present results. 
This is because the underlying potential was adjusted for each nuclide so as to reproduce the shift and width of the corresponding antiprotonic atom~\cite{Batty1983-sn,Batty2001-hh}, whereas we employ the common form of the optical potential with a single $b_0$. 

\section{Isovector term and realistic density distributions}\label{sec:isovector_potential}

Thus far, we have assumed the 2pF nucleon density distributions and investigated the general features of the scattering length.
Inspired by a recent paper by Yoshimura et al.~\cite{Yoshimura2025-vt}, we consider more realistic nucleon density distributions computed using density functional theory (DFT)~\cite{Hohenberg1964,Kohn1965,Vautherin1972,Kohn1999}.
They calculated the shifts and widths of the $5g$ and $6h$ states in antiprotonic \Ca{40} and \Ca{48} atoms with several optical potentials, including $b_0=1.3+1.9i\,\mathrm{fm}$, $b_1=0$ (Type-I), which corresponds to the Friedman2005 potential, and $b_0=1.3+1.9i\,\mathrm{fm}$, $b_1=-8.0+1.7i\,\mathrm{fm}$ (Type-II).
While the Type-I potential with the isoscalar $b_0$ term alone fails to account for the measured large shift of \Ca{48}, the Type-II potential reproduces it considerably better.
This led them to argue that additional terms beyond the conventional isoscalar one, Eq.~(\ref{eq:optical_potential_b0only}) --- the isovector term and/or the $p$-wave term --- are required, at least for the calcium isotopes.
They further found that, once the isovector term is included, the calculated quantities depend considerably on the adopted density profile.
Motivated by this finding, we present the scattering length for both the Type-I and Type-II potentials with DFT-based density distributions in this section.
We use the same set of density distributions as in Ref.~\cite{Yoshimura2025-vt}, where they were obtained from spherical Hartree--Fock--Bogoliubov calculations~\cite{Dobaczewski1984} with several Skyrme parametrizations: SLy4 and SLy5~\cite{Chabanat1998}, SkM$^*$~\cite{Bartel1982}, SAMi~\cite{RocaMaza2012}, SGII~\cite{VanGiai1981}, \mbox{UNEDF0}~\cite{Kortelainen2010}, \mbox{UNEDF1}~\cite{Kortelainen2012}, \mbox{UNEDF2}~\cite{Kortelainen2014}, and HFB9~\cite{Goriely2005}.

\begin{figure}[t!]
  \centering
  \includegraphics[width=\columnwidth]{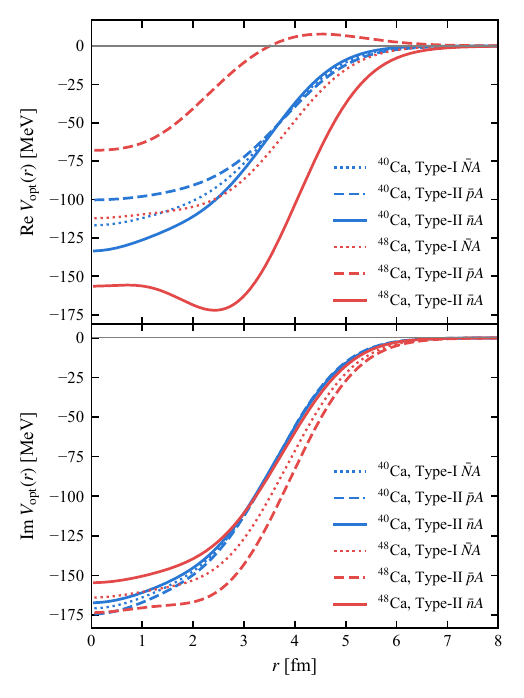}
  \caption{Radial shape of $\operatorname{Re}V_{\mathrm{opt}}(r)$ (upper) and
$\operatorname{Im}V_{\mathrm{opt}}(r)$ (lower) for the three types of potentials
--- \TypeINA (dotted), \TypeIIpA (dashed), and \TypeIInA (solid) ---
for \Ca{40} (blue) and \Ca{48} (red), using the SLy4 parametrization.}\label{fig:a_iscovector_Vopt}
\end{figure}

From Eq.~(\ref{eq:optical_potential_antineutron}), the Type-I potential is common to the $\bar{p}A$ and $\bar{n}A$ interactions.
On the other hand, the Type-II potential results in opposite signs for the isovector term, represented as Eqs.~(\ref{eq:optical_potential_full}) and (\ref{eq:optical_potential_antineutron}), for these two interactions.
To avoid confusion, we refer to these three optical potentials with common $b_0$ and $b_1$ parameters as ``\TypeINA'', ``\TypeIIpA'', and ``\TypeIInA'' potentials, respectively.
As an example, these optical potentials with the nucleon density distribution under the SLy4 parametrization are displayed in Fig.~\ref{fig:a_iscovector_Vopt} for \Ca{40} and \Ca{48}.
For \Ca{48}, where $\delta\rho(r)>0$, the combination $\operatorname{Re}b_1<0$ and $\operatorname{Im}b_1>0$ makes the \TypeIInA interaction more attractive and slightly less absorptive than the \TypeIIpA interaction.
On the other hand, \Ca{40} satisfies $\rho_n(r)\approx \rho_p(r)$; hence, the difference among the three potentials is less significant.

\begin{figure}[t!]
  \centering
  \includegraphics[width=\columnwidth]{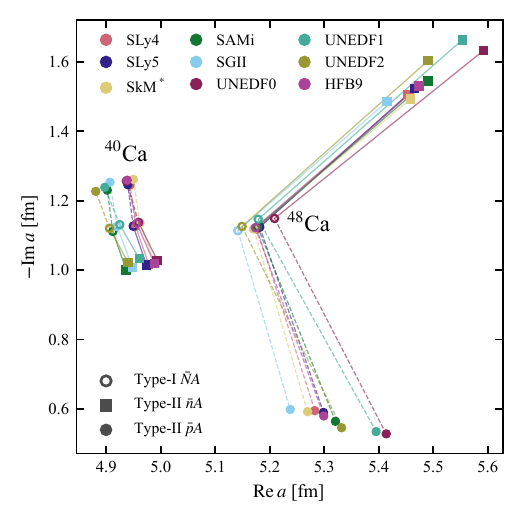}
  \caption{Strong-interaction scattering lengths for \Ca{40} and \Ca{48} with the Type-I potential (open circles), and with the Type-II potential applied to the $\bar{n}A$ system (filled squares), and to the $\bar{p}A$ system (filled circles). Different colors correspond to different Skyrme parametrizations, labeled in the legend.}\label{fig:a_isovector}
\end{figure}

The scattering lengths for these optical potentials are plotted in
Fig.~\ref{fig:a_isovector} for \Ca{40} and \Ca{48}.
The spread of $\operatorname{Re}a$ and $-\operatorname{Im}a$ for each potential over the nine Skyrme parametrizations is as follows.
For the \TypeINA (\mbox{Friedman2005}) potential, the spreads due to the density distributions are within $\pm (1\text{--}2)\%$, for both \Ca{40} and \Ca{48}, whereas varying either the real or the imaginary part of $b_0$
by $\pm0.1\,\mathrm{fm}$ around the central value in
Eq.~(\ref{eq:optical_potential_folded}), with the other component and the density distributions held fixed, changes each component of the scattering length by about $0.015\,\mathrm{fm}$.
Inclusion of the isovector terms in \TypeIInA and \TypeIIpA modifies the scattering length.
Whereas the isovector-induced difference between \TypeINA and \TypeIInA in \Ca{40} is small ($\varDelta(\operatorname{Re}a)=0.024\text{--}0.036\,\mathrm{fm}$ and $\varDelta(-\operatorname{Im}a)=-(0.098\text{--}0.118)\,\mathrm{fm}$), the isovector-induced difference in \Ca{48} is very large, $\varDelta(\operatorname{Re}a)=0.274\text{--}0.383\,\mathrm{fm}$, and $\varDelta(-\operatorname{Im}a)=0.372\text{--}0.517\,\mathrm{fm}$, reflecting the large difference between the two potentials shown in Fig.~\ref{fig:a_iscovector_Vopt}.
Moreover, the spread itself is magnified by the isovector term.
Although the isovector term enters \TypeIInA (Eq.~(\ref{eq:optical_potential_antineutron})) and \TypeIIpA (Eq.~(\ref{eq:optical_potential_full})) with opposite signs, this does not simply move the scattering length in opposite directions.
The real part of the scattering length is increased for both the \TypeIInA and \TypeIIpA potentials for \Ca{48}, which could be interpreted as a large modification of the optical potential shown in Fig.~\ref{fig:a_iscovector_Vopt} due to the isovector term with a large $|b_1|$.

\begin{figure}[ht!]
  \centering
  \includegraphics[width=\columnwidth]{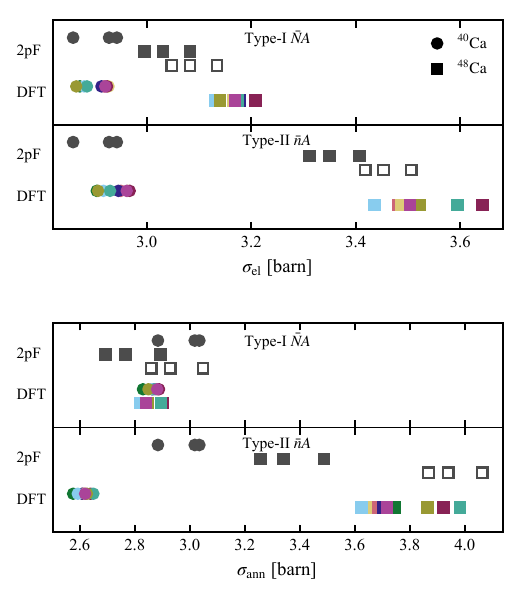}
  \caption{Elastic and annihilation $s$-wave cross sections, $\sigma_\mathrm{el}$ and $\sigma_\mathrm{ann}$, for antineutron--Ca scattering (top: \TypeINA; bottom: \TypeIInA), at $p_{\bar{n}}=9\,\mathrm{MeV}/c$, calculated using the same scattering lengths as in Fig.~\ref{fig:a_isovector}. The definition of the colored circles (\Ca{40}) and squares (\Ca{48}) is the same as in Fig.~\ref{fig:a_isovector}. For comparison, the cross sections for 2pF nucleon density distributions are also shown: filled circles (\Ca{40}), filled squares (\Ca{48}, neutron skin), and open squares (\Ca{48}, neutron halo) with three different sets of $c_p$ and $z_p$.}\label{fig:a_isovector_xsec}
\end{figure}

Finally, using the above scattering lengths, we calculated the elastic and annihilation $s$-wave cross sections for antineutron--Ca scattering, $\sigma_\mathrm{el}$ (Eq.~(\ref{eq:sigma_el})) and $\sigma_\mathrm{ann}$ (Eq.~(\ref{eq:sigma_ann})), at an incident antineutron momentum of $9\,\mathrm{MeV}/c$.
The results are summarized in Fig.~\ref{fig:a_isovector_xsec}.
For comparison, the same calculation was performed assuming a 2pF distribution for $\rho_p(r)$ for both \Ca{40} and \Ca{48}, using three different sets of $c_p$ and $z_p$ tabulated in Ref.~\cite{Wu2025-gi}; for \Ca{40}, $\rho_n(r)=\rho_p(r)$ is assumed, while for \Ca{48}, skin-type and halo-type neutron distributions are assumed instead (see Sec.~\ref{sec:mass_no_dependence}).
For \Ca{40} with \TypeINA, the cross sections are rather insensitive to the density distribution.
In contrast, the annihilation cross section for \TypeIInA is sensitive to $\delta\rho(r)$, which is fixed to be zero in the 2pF densities.
The cross sections for \Ca{48} spread more widely due to the density distributions, particularly in the case of \TypeIInA.
This indicates that these cross sections are sensitive to a finite isovector term; turning this sensitivity into a quantitative determination of $b_1$ will require precise knowledge of the nucleon density distributions.

\section{Summary and outlook}\label{sec:summary}
Antineutron--nucleus scattering lengths are expected to be independently determined through low-energy antineutron scattering experiments in the $s$-wave regime.
For nuclei beyond the few-body systems considered in \textit{ab initio} calculations based on $\bar{N}N$ interactions, optical potentials constrained by antiprotonic-atom data provide model-based estimates of antineutron scattering lengths.
We therefore studied the antineutron--nucleus scattering length based on these optical potentials to prepare for such measurements.

First, an analytic approximation of the scattering length, $a_\text{approx}$, for a simple 2pF potential with a complex potential depth $-U_0$ and geometrical parameters of the potential (the half-density radius $c$ and the diffuseness $z$) was obtained.
For fixed $c$ and $z$, $\operatorname{Re}a_\text{approx}-c$ and $\operatorname{Im}a_\text{approx}$ depend primarily on $|U_0|$ and $\mathrm{arg}\,U_0$, respectively.
Measurements of both components would therefore constrain the complex potential depth within this parametrization.
Moreover, the dependence on $z$ highlights the importance of the potential shape in the surface region.

Next, the mass-number dependence of the scattering length for an isoscalar potential proportional to $\rho(r)=\rho_n(r)+\rho_p(r)$ was investigated.
Both $\rho_p(r)$ and $\rho_n(r)$ were assumed to follow 2pF distributions. For a heavy nuclide with $N>Z$, different assumptions on $\rho_n(r)$, much less constrained experimentally than $\rho_p(r)$, resulted in a sizable change of the scattering length.
The magnitude of the imaginary part, $-\operatorname{Im}a$, is larger for the finite-range Friedman2005 potential than for the zero-range Batty1995 potential for each assumption on the density distribution.

We also investigated the impact of an isovector extension of the optical potential as well as realistic nucleon density distributions from DFT calculations for \Ca{40} and \Ca{48}.
In particular, the isovector term with $b_1=-8.0+1.7i\,\mathrm{fm}$ increases $\operatorname{Re}a$ by $0.3\text{--}0.4\,\mathrm{fm}$ and $-\operatorname{Im}a$ by $0.4\text{--}0.5\,\mathrm{fm}$ for \Ca{48}, while the spread among different nucleon density distributions is magnified by the isovector term.

Taken together, these calculations quantify the dependence of the predicted scattering lengths on the adopted density distributions and optical-potential parametrizations.
Measurements of both low-energy elastic and annihilation cross sections would provide independent constraints on the threshold interaction and test the optical-potential model.

We propose a two-stage strategy to test and refine the optical-potential approach using the scattering length determined directly by antineutron scattering experiments.
As shown in Sec.~\ref{sec:isovector_potential}, the optical potential without the isovector term, Eq.~(\ref{eq:optical_potential_b0only}), is a reasonable approximation for $N=Z$ nuclides at the first stage,
because the possible contribution of the isovector term with a small $\delta\rho(r)$ would be minimized.

\begin{enumerate}
\item A measurement of the antineutron scattering length for an $N=Z$ nuclide, such as $^{12}\mathrm{C}$, $^{28}\mathrm{Si}$, and \Ca{40}, can be used to test the ansatz of the optical potential, Eq.~(\ref{eq:optical_potential_b0only}).
The key question is whether a common (global) $b_0$ parameter for a fixed interaction range $\beta$ can account for the experimental scattering lengths for various $N=Z$ nuclides.

\item The measurement can be extended to $N>Z$ nuclides.
As the role of the isovector term becomes more significant, the quantitative evaluation of the $b_1$ parameter requires realistic neutron density distributions, e.g., from DFT calculations, \textit{ab initio} calculations~\cite{Papenbrock2024-pi}, or proton elastic scattering~\cite{Terashima2008-ky,Zenihiro2010-gk,Zenihiro2018-rt}.
\end{enumerate}

The first direct determination of the antineutron--nucleus scattering length would serve as a critical check of the optical potential in the low-density approximation, which accounts for antiprotonic-atom data globally.
At the same time, it would also provide independent information on the antinucleon--nucleus strong interaction at zero energy.


\appendix
\setcounter{figure}{0}

\section{Derivation of the analytic approximation to the scattering length for a 2pF potential}\label{sec:scattering_length_2pF}

\begin{figure*}[t!]
  \centering
  \includegraphics[width=\textwidth]{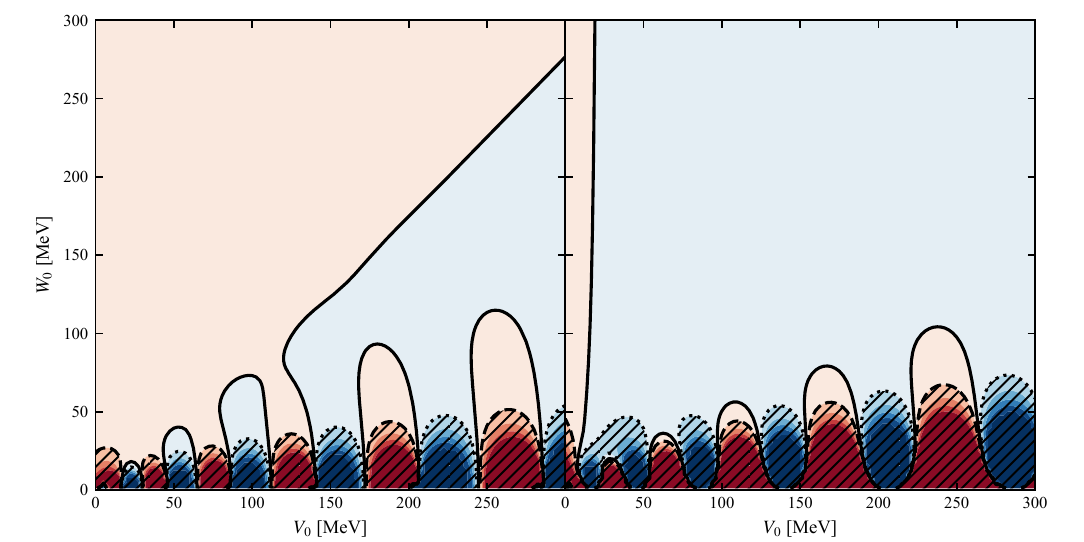}
  \caption{Accuracy of the analytic approximation. Deviation of the scattering length $a_\mathrm{approx}$ approximated by Eq.~(\ref{eq:approx_scattering_length}) from the numerically calculated one shown in Fig.~\ref{fig:scattering_length_2pF}, for a potential depth of $-(V_0+iW_0)$, with the half-density radius and the diffuseness fixed to $5\,\mathrm{fm}$ and $0.5\,\mathrm{fm}$, respectively, in units of \%: real part (left) and imaginary part (right). The solid, dashed, and dotted lines represent $0\%$, $+1\%$, and $-1\%$ deviation, respectively. The deviation exceeds 1\% in magnitude in the hatched region.}\label{fig:approx_deviation_2pF}
\end{figure*}

The exact expression for the scattering length $a$ for a 2pF potential $V(r)=-U_0/(1+\exp[(r-c)/z])$ is given in Ref.~\cite{Liverts2021}.
Following the notation of Ref.~\cite{Liverts2021}, we define:
\begin{gather}
  K=\frac{\sqrt{2\mu U_0}}{\hbar},\quad \alpha=\frac{z}{\hbar}\sqrt{2\mu U_0}=zK,\\
  F(\alpha,x) = x^{i\alpha}{}_2F_1(i\alpha,i\alpha; 1 + 2i\alpha; -x),
\end{gather}
where ${}_2F_1(a,b;c;z)$ is the Gauss hypergeometric function:
\begin{gather}
  {}_2F_1(a,b;c;z)=\sum_{n=0}^{\infty}\frac{(a)_n(b)_n}{(c)_n}\frac{z^n}{n!},
\intertext{with}
  (a)_n=\frac{\Gamma(a+n)}{\Gamma(a)}=a(a+1)\cdots(a+n-1).
\end{gather}
Then, the scattering length $a$ is given by
\begin{gather}
  a=c+z\left[2\gamma+\psi(i\alpha)+\psi(-i\alpha)-i\pi\coth (\pi\alpha)\frac{P+Q}{P-Q}\right],\label{eq:scattering-length-2pF}
\intertext{with}
  P=F(\alpha,x_0)\Gamma^2(i\alpha)\Gamma(-2i\alpha),\\
  Q=F(-\alpha,x_0)\Gamma^2(-i\alpha)\Gamma(2i\alpha),\\
  x_0=\exp(-c/z),
\end{gather}
where $\psi$ is the digamma function and $\gamma=0.577\ldots$ is Euler's constant.
The original expression in Ref.~\cite{Liverts2021} applies to a real potential with $\alpha$ real; the present form extends it to a complex potential.

We now derive the asymptotic form of the scattering length for a 2pF potential in the limit of large $c$ at fixed $z$ and $U_0$.

Since
\begin{align}
  x_0^{\pm i\alpha}=\exp(\mp i\alpha c/z)=\exp(\mp iKc),
\end{align}
and
\begin{align}
  {}_2F_1(\pm i\alpha,\pm i\alpha; 1 \pm 2i\alpha; -x_0)=1+\mathcal{O}(x_0),
\end{align}
we obtain
\begin{align}
F(\pm \alpha ,x_0)=\exp(\mp iKc)[1+\mathcal{O}(x_0)]\approx \exp(\mp iKc),
\end{align}
as $x_0=\exp(-c/z)$ becomes exponentially small for large $c$. 
Then, the ratio $Q/P$ can be approximated as
\begin{align}
  \frac{Q}{P}\approx \exp(2iKc)\frac{\Gamma^2(-i\alpha)\Gamma(2i\alpha)}{\Gamma^2(i\alpha)\Gamma(-2i\alpha)}.
\end{align}
Since  all factors other than $\exp(2iKc)$ remain finite, the modulus $|Q/P|$ behaves as $\exp[-2(\operatorname{Im}K)c]$, decaying to zero for sufficiently large $c$.
Therefore, the scattering length $a$ in Eq.~(\ref{eq:scattering-length-2pF}) is given by
\begin{align}
a \approx c+z\left[2\gamma+\psi(i\alpha)+\psi(-i\alpha)-i\pi\coth (\pi\alpha)\right].
\end{align}
Defining $\zeta\equiv -i\alpha=-izK$, we can rewrite the above expression as
\begin{align}
a &\approx c+z\left[2\gamma +\psi(\zeta)+\psi(\zeta+1) \right].\label{eq:scattering-length-Rinf}\\
&= c+2z\left[\gamma+\psi(\zeta)+\frac{1}{2\zeta}\right].\label{eq:scattering-length-Rinf2}
\end{align}
This follows from the reflection formula:
\begin{align}
\psi(1-(-\zeta))-\psi(-\zeta)=\pi\cot(-\pi\zeta)=-i\pi\coth (\pi\alpha)
\end{align}
and the recurrence formula
$\psi(\zeta+1)=\psi(\zeta)+1/\zeta$.
Here the principal square root is taken, so that $\zeta$ lies in the fourth quadrant and  $\psi(\zeta)$ is analytic.

Since the digamma function $\psi(\zeta)$ has the asymptotic expansion
\begin{align} 
  \psi(\zeta)=\ln \zeta-\frac{1}{2\zeta}-\frac{1}{12\zeta^2}+\cdots,\label{eq:asymptotic}
\end{align}
the scattering length (\ref{eq:scattering-length-Rinf2}) can be approximated as
\begin{align}
a\approx c+2\gamma z+2z\ln \zeta-\frac{z}{6\zeta^2},\label{eq:scattering-length-Rinf3}
\end{align}
where $\ln\zeta$ is the principal logarithm, as given in Eq.~(\ref{eq:logzeta}).

Note that the asymptotic expansion breaks down in the limit of $z\to 0$ at fixed $c$ and $K$ corresponding to a square-well potential.
As $\psi(\zeta)\to -1/\zeta-\gamma$ in this limit, the r.h.s. of Eq.~(\ref{eq:scattering-length-Rinf2}) reduces to the strong-absorption square-well expression, $c-i/K$, under the condition $\mathrm{Im}(K)c\gg1$.

\section*{Acknowledgements}
The authors are grateful to T.~Naito for providing the nucleon density distributions calculated using DFT.
H.F. acknowledges useful discussions with D.~Jido and K.~Yoshimura.

\section*{Funding}
This work was supported by ASUNARO Grant from Institute of Science Tokyo, by Yamada Science Foundation, and by JSPS KAKENHI Grant Number JP26K00721.

\section*{Data availability}
The data are available from the corresponding author upon request.

\section*{Declaration of generative AI and AI-assisted technologies in the manuscript preparation process}
During the preparation of this work, the authors used Claude (Anthropic) and ChatGPT (OpenAI) to assist with English language editing, the development and verification of Python numerical codes used in the calculations, and algebraic manipulation in the derivation presented in \ref{sec:scattering_length_2pF}.
After using these tools, the authors reviewed and edited the content as needed and take full responsibility for the content of the published article.

\bibliographystyle{elsarticle-num} 
\bibliography{reference}

\begin{thebibliography}{10}
\expandafter\ifx\csname url\endcsname\relax
  \def\url#1{\texttt{#1}}\fi
\expandafter\ifx\csname urlprefix\endcsname\relax\def\urlprefix{URL }\fi
\expandafter\ifx\csname href\endcsname\relax
  \def\href#1#2{#2} \def\path#1{#1}\fi

\bibitem{Sears1992}
V.~F. Sears, Neutron scattering lengths and cross sections, Neutron News 3~(3) (1992) 26--37.
\newblock \href {https://doi.org/10.1080/10448639208218770} {\path{doi:10.1080/10448639208218770}}.

\bibitem{Deser1954-sq}
S.~Deser, M.~L. Goldberger, K.~Baumann, W.~Thirring, Energy level displacements in pi-mesonic atoms, Phys. Rev. 96 (1954) 774--776.
\newblock \href {https://doi.org/10.1103/physrev.96.774} {\path{doi:10.1103/physrev.96.774}}.

\bibitem{Trueman1961-aq}
T.~L. Trueman, Energy level shifts in atomic states of strongly-interacting particles, Nucl. Phys. 26 (1961) 57--67.
\newblock \href {https://doi.org/10.1016/0029-5582(61)90115-8} {\path{doi:10.1016/0029-5582(61)90115-8}}.

\bibitem{Filippi2025}
A.~Filippi, H.~Fujioka, T.~Higuchi, L.~Venturelli, \href{https://arxiv.org/abs/2503.06972}{Novel concept for low-energy antineutron production and its application for antineutron scattering experiments} (2025).
\newblock \href {http://arxiv.org/abs/2503.06972} {\path{arXiv:2503.06972}}, \href {https://doi.org/10.48550/arXiv.2503.06972} {\path{doi:10.48550/arXiv.2503.06972}}.
\newline\urlprefix\url{https://arxiv.org/abs/2503.06972}

\bibitem{Amsler:2930906}
C.~Amsler, D.~Calvo, A.~Feliciello, A.~Filippi, H.~Fujioka, T.~Higuchi, L.~Venturelli, \href{https://cds.cern.ch/record/2930906}{Low-energy antineutron beamline at the antiproton decelerator for scattering experiments}, Tech. Rep. CERN-SPSC-2025-010, SPSC-I-261, CERN, Geneva (2025).
\newline\urlprefix\url{https://cds.cern.ch/record/2930906}

\bibitem{Filippi2026-hs}
A.~Filippi, \href{https://arxiv.org/abs/2601.11390}{The unfinished picture of low-energy antineutron interactions: open issues and hints for future research possibilities} (2026).
\newblock \href {http://arxiv.org/abs/2601.11390} {\path{arXiv:2601.11390}}, \href {https://doi.org/10.48550/arXiv.2601.11390} {\path{doi:10.48550/arXiv.2601.11390}}.
\newline\urlprefix\url{https://arxiv.org/abs/2601.11390}

\bibitem{Astrua2002-ez}
M.~Astrua, E.~Botta, T.~Bressani, D.~Calvo, C.~Casalegno, A.~Feliciello, A.~Filippi, S.~Marcello, M.~Agnello, F.~Iazzi, Antineutron--nucleus annihilation cross sections below 400 {MeV}/c, Nucl. Phys. A 697 (2002) 209--224.
\newblock \href {https://doi.org/10.1016/s0375-9474(01)01252-0} {\path{doi:10.1016/s0375-9474(01)01252-0}}.

\bibitem{Carbonell2023-lt}
J.~Carbonell, G.~Hupin, S.~Wycech, Comparison of {$\bar{N}N$} optical models, Eur. Phys. J. A 59 (2023) 259.
\newblock \href {https://doi.org/10.1140/epja/s10050-023-01161-x} {\path{doi:10.1140/epja/s10050-023-01161-x}}.

\bibitem{Gotta2004-bq}
D.~Gotta, Precision spectroscopy of light exotic atoms, Prog. Part. Nucl. Phys. 52 (2004) 133--195.
\newblock \href {https://doi.org/10.1016/j.ppnp.2003.09.003} {\path{doi:10.1016/j.ppnp.2003.09.003}}.

\bibitem{Lazauskas2021-wp}
R.~Lazauskas, J.~Carbonell, Antiproton-deuteron hydrogenic states in optical models, Phys. Lett. B 820 (2021) 136573.
\newblock \href {https://doi.org/10.1016/j.physletb.2021.136573} {\path{doi:10.1016/j.physletb.2021.136573}}.

\bibitem{Duerinck2023-qw}
P.-Y. Duerinck, R.~Lazauskas, J.~Dohet-Eraly, Antiproton-deuteron hydrogenic states from a coupled-channel approach, Phys. Rev. C 108 (2023) 054003.
\newblock \href {https://doi.org/10.1103/physrevc.108.054003} {\path{doi:10.1103/physrevc.108.054003}}.

\bibitem{Duerinck2026-mv}
P.-Y. Duerinck, R.~Lazauskas, \textit{Ab initio} description of {$\bar{p}+{}^3\mathrm{H}$} and {$\bar{p}+{}^3\mathrm{He}$} systems in optical models, Phys. Rev. C 113 (2026) 054003.
\newblock \href {https://doi.org/10.1103/c7s6-x4v9} {\path{doi:10.1103/c7s6-x4v9}}.

\bibitem{Dehghani2026-sk}
A.~Dehghani, G.~Hupin, S.~Quaglioni, P.~Navr^^c3^^a1til, Light antiproton-nucleus systems at low energies with the \textit{ab initio} {NCSM}/{RGM} method, Phys. Rev. C 114 (2026) 014613.
\newblock \href {https://doi.org/10.1103/ldr8-xfp1} {\path{doi:10.1103/ldr8-xfp1}}.

\bibitem{Dehghani2026-pq}
A.~Dehghani, G.~Hupin, S.~Quaglioni, P.~Navr{\'a}til, \href{https://arxiv.org/abs/2608.16552}{Ab initio evidence for surface-dominated antiproton annihilation in ${}^4\mathrm{He}$} (2026).
\newblock \href {http://arxiv.org/abs/2608.16552} {\path{arXiv:2608.16552}}, \href {https://doi.org/10.48550/arXiv.2608.16552} {\path{doi:10.48550/arXiv.2608.16552}}.
\newline\urlprefix\url{https://arxiv.org/abs/2608.16552}

\bibitem{Batty1997-nd}
C.~J. Batty, E.~Friedman, A.~Gal, Strong interaction physics from hadronic atoms, Phys. Rep. 287 (1997) 385--445.
\newblock \href {https://doi.org/10.1016/s0370-1573(97)00011-2} {\path{doi:10.1016/s0370-1573(97)00011-2}}.

\bibitem{Friedman2007-vu}
E.~Friedman, A.~Gal, In-medium nuclear interactions of low-energy hadrons, Phys. Rep. 452 (2007) 89--153.
\newblock \href {https://doi.org/10.1016/j.physrep.2007.08.002} {\path{doi:10.1016/j.physrep.2007.08.002}}.

\bibitem{Batty1995-cb}
C.~J. Batty, E.~Friedman, A.~Gal, Density-dependent {$\bar{p}$}-nucleus optical potentials from global fits to $\bar{p}$ atom data, Nucl. Phys. A 592 (1995) 487--512.
\newblock \href {https://doi.org/10.1016/0375-9474(95)00308-N} {\path{doi:10.1016/0375-9474(95)00308-N}}.

\bibitem{Friedman2005-yl}
E.~Friedman, A.~Gal, J.~Mare\v{s}, Antiproton-nucleus potentials from global fits to antiprotonic {X}-rays and radiochemical data, Nucl. Phys. A 761 (2005) 283--295.
\newblock \href {https://doi.org/10.1016/j.nuclphysa.2005.08.001} {\path{doi:10.1016/j.nuclphysa.2005.08.001}}.

\bibitem{Yoshimura2025-vt}
K.~Yoshimura, S.~Yasunaga, D.~Jido, J.~Yamagata-Sekihara, S.~Hirenzaki, Interrelation between {$\bar{p}$}-{Ca} atom spectra and nuclear density profiles, Prog. Theor. Exp. Phys. 2025 (2025) 113D02.
\newblock \href {https://doi.org/10.1093/ptep/ptaf140} {\path{doi:10.1093/ptep/ptaf140}}.

\bibitem{Trzcinska2001-mp}
A.~Trzci^^c5^^84ska, J.~Jastrz^^c4^^99bski, T.~Czosnyka, T.~von Egidy, K.~Gulda, F.~J. Hartmann, J.~Iwanicki, B.~Ketzer, M.~Kisieli^^c5^^84ski, B.~K^^c5^^82os, W.~Kurcewicz, P.~Lubi^^c5^^84ski, P.~J. Napiorkowski, L.~Pie^^c5^^84kowski, R.~Schmidt, E.~Widmann, Information on antiprotonic atoms and the nuclear periphery from the {PS209} experiment, Nucl. Phys. A 692 (2001) 176--181.
\newblock \href {https://doi.org/10.1016/S0375-9474(01)01176-9} {\path{doi:10.1016/S0375-9474(01)01176-9}}.

\bibitem{Friedman2008-sf}
E.~Friedman, A.~Gal, Realistic calculations of nuclear disappearance lifetimes induced by {$n\bar{n}$} oscillations, Phys. Rev. D 78 (2008) 016002.
\newblock \href {https://doi.org/10.1103/PhysRevD.78.016002} {\path{doi:10.1103/PhysRevD.78.016002}}.

\bibitem{Abe2021-td}
{{K. Abe et al. [Super-Kamiokande Collaboration]}}, {Neutron-antineutron oscillation search using a 0.37 megaton-years exposure of Super-Kamiokande}, Phys. Rev. D 103 (2021) 012008.
\newblock \href {https://doi.org/10.1103/PhysRevD.103.012008} {\path{doi:10.1103/PhysRevD.103.012008}}.

\bibitem{Trzcinska2001-fn}
A.~Trzci^^c5^^84ska, J.~Jastrz^^c4^^99bski, P.~Lubi^^c5^^84ski, F.~J. Hartmann, R.~Schmidt, T.~von Egidy, B.~K^^c5^^82os, Neutron density distributions deduced from antiprotonic atoms, Phys. Rev. Lett. 87 (2001) 082501.
\newblock \href {https://doi.org/10.1103/PhysRevLett.87.082501} {\path{doi:10.1103/PhysRevLett.87.082501}}.

\bibitem{Aumann2022-qk}
T.~Aumann, et~al., {PUMA}, {antiProton} unstable matter annihilation, Eur. Phys. J. A 58 (2022) 88.
\newblock \href {https://doi.org/10.1140/epja/s10050-022-00713-x} {\path{doi:10.1140/epja/s10050-022-00713-x}}.

\bibitem{Batty2001-hh}
C.~J. Batty, E.~Friedman, A.~Gal, Unified optical-model approach to low-energy antiproton annihilation on nuclei and to antiprotonic atoms, Nucl. Phys. A 689 (2001) 721--740.
\newblock \href {https://doi.org/10.1016/S0375-9474(00)00608-4} {\path{doi:10.1016/S0375-9474(00)00608-4}}.

\bibitem{Batty1983-sn}
C.~J. Batty, Hadron-nucleus scattering lengths derived from exotic atom data, Nucl. Phys. A 411 (1983) 399--416.
\newblock \href {https://doi.org/10.1016/0375-9474(83)90538-9} {\path{doi:10.1016/0375-9474(83)90538-9}}.

\bibitem{Nesvizhevsky2019-vw}
V.~V. Nesvizhevsky, V.~Gudkov, K.~V. Protasov, W.~M. Snow, A.~Y. Voronin, Experimental approach to search for free neutron-antineutron oscillations based on coherent neutron and antineutron mirror reflection, Phys. Rev. Lett. 122 (2019) 221802.
\newblock \href {https://doi.org/10.1103/PhysRevLett.122.221802} {\path{doi:10.1103/PhysRevLett.122.221802}}.

\bibitem{Gudkov2020-dn}
V.~Gudkov, V.~V. Nesvizhevsky, K.~V. Protasov, W.~M. Snow, A.~Y. Voronin, A new approach to search for free neutron-antineutron oscillations using coherent neutron propagation in gas, Phys. Lett. B 808 (2020) 135636.
\newblock \href {https://doi.org/10.1016/j.physletb.2020.135636} {\path{doi:10.1016/j.physletb.2020.135636}}.

\bibitem{Protasov2020-gy}
K.~V. Protasov, V.~Gudkov, E.~A. Kupriyanova, V.~V. Nesvizhevsky, W.~M. Snow, A.~Y. Voronin, Theoretical analysis of antineutron-nucleus data needed for antineutron mirrors in neutron-antineutron oscillation experiments, Phys. Rev. D 102 (2020) 075025.
\newblock \href {https://doi.org/10.1103/PhysRevD.102.075025} {\path{doi:10.1103/PhysRevD.102.075025}}.

\bibitem{Shima2025-ie}
T.~Shima, Experimental search for neutron^^e2^^80^^93antineutron oscillation with the use of ultra-cold neutrons revisited, Symmetry 17 (2025) 1524.
\newblock \href {https://doi.org/10.3390/sym17091524} {\path{doi:10.3390/sym17091524}}.

\bibitem{Fujioka2026-qp}
H.~Fujioka, T.~Higuchi, Impact of a reflecting material on a search for neutron^^e2^^80^^93antineutron oscillations using ultracold neutrons, Prog. Theor. Exp. Phys. 2026 (2026) 023C01.
\newblock \href {https://doi.org/10.1093/ptep/ptaf189} {\path{doi:10.1093/ptep/ptaf189}}.

\bibitem{Calogero1967}
F.~Calogero, Variable Phase Approach to Potential Scattering, Mathematics in Science and Engineering, Volume 35, Academic Press, New York, 1967.

\bibitem{Karmanov2000-tm}
V.~A. Karmanov, K.~V. Protasov, A.~Y. Voronin, Antiproton-deuteron annihilation at low energies, Eur. Phys. J. A 8 (2000) 429--434.
\newblock \href {https://doi.org/10.1007/s100500070098} {\path{doi:10.1007/s100500070098}}.

\bibitem{Wu2025-gi}
T.~Y. Wu, B.~H. Sun, H.~H. Xie, J.~Y. Xu, G.~Guo, Point-proton density distributions of stable nuclei, At. Data Nucl. Data Tables 165 (2025) 101733.
\newblock \href {https://doi.org/10.1016/j.adt.2025.101733} {\path{doi:10.1016/j.adt.2025.101733}}.

\bibitem{Vinas2014-aw}
X.~Vi^^c3^^b1as, M.~Centelles, X.~Roca-Maza, M.~Warda, Density dependence of the symmetry energy from neutron skin thickness in finite nuclei, Eur. Phys. J. A 50 (2014) 27.
\newblock \href {https://doi.org/10.1140/epja/i2014-14027-8} {\path{doi:10.1140/epja/i2014-14027-8}}.

\bibitem{Hohenberg1964}
P.~Hohenberg, W.~Kohn, Inhomogeneous electron gas, Phys. Rev. 136 (1964) B864--B871.
\newblock \href {https://doi.org/10.1103/PhysRev.136.B864} {\path{doi:10.1103/PhysRev.136.B864}}.

\bibitem{Kohn1965}
W.~Kohn, L.~J. Sham, Self-consistent equations including exchange and correlation effects, Phys. Rev. 140 (1965) A1133--A1138.
\newblock \href {https://doi.org/10.1103/PhysRev.140.A1133} {\path{doi:10.1103/PhysRev.140.A1133}}.

\bibitem{Vautherin1972}
D.~Vautherin, D.~M. Brink, Hartree-{F}ock calculations with {S}kyrme's interaction. {I}. spherical nuclei, Phys. Rev. C 5 (1972) 626--647.
\newblock \href {https://doi.org/10.1103/PhysRevC.5.626} {\path{doi:10.1103/PhysRevC.5.626}}.

\bibitem{Kohn1999}
W.~Kohn, Nobel lecture: Electronic structure of matter---wave functions and density functionals, Rev. Mod. Phys. 71 (1999) 1253--1266.
\newblock \href {https://doi.org/10.1103/RevModPhys.71.1253} {\path{doi:10.1103/RevModPhys.71.1253}}.

\bibitem{Dobaczewski1984}
J.~Dobaczewski, H.~Flocard, J.~Treiner, {H}artree--{F}ock--{B}ogolyubov description of nuclei near the neutron-drip line, Nucl. Phys. A 422 (1984) 103--139.
\newblock \href {https://doi.org/10.1016/0375-9474(84)90433-0} {\path{doi:10.1016/0375-9474(84)90433-0}}.

\bibitem{Chabanat1998}
E.~Chabanat, P.~Bonche, P.~Haensel, J.~Meyer, R.~Schaeffer, A {S}kyrme parametrization from subnuclear to neutron star densities {P}art {II}. {N}uclei far from stabilities, Nucl. Phys. A 635 (1998) 231--256, [Erratum: Nucl. Phys. A 643 (1998) 441].
\newblock \href {https://doi.org/10.1016/S0375-9474(98)00180-8} {\path{doi:10.1016/S0375-9474(98)00180-8}}.

\bibitem{Bartel1982}
J.~Bartel, P.~Quentin, M.~Brack, C.~Guet, H.~B. H{\aa}kansson, Towards a better parametrisation of {S}kyrme-like effective forces: {A} critical study of the {S}k{M} force, Nucl. Phys. A 386 (1982) 79--100.
\newblock \href {https://doi.org/10.1016/0375-9474(82)90403-1} {\path{doi:10.1016/0375-9474(82)90403-1}}.

\bibitem{RocaMaza2012}
X.~Roca-Maza, G.~Col{\`o}, H.~Sagawa, New {S}kyrme interaction with improved spin-isospin properties, Phys. Rev. C 86 (2012) 031306.
\newblock \href {https://doi.org/10.1103/PhysRevC.86.031306} {\path{doi:10.1103/PhysRevC.86.031306}}.

\bibitem{VanGiai1981}
N.~{Van Giai}, H.~Sagawa, Spin-isospin and pairing properties of modified {S}kyrme interactions, Phys. Lett. B 106 (1981) 379--382.
\newblock \href {https://doi.org/10.1016/0370-2693(81)90646-8} {\path{doi:10.1016/0370-2693(81)90646-8}}.

\bibitem{Kortelainen2010}
M.~Kortelainen, T.~Lesinski, J.~Mor{\'e}, W.~Nazarewicz, J.~Sarich, N.~Schunck, M.~V. Stoitsov, S.~Wild, Nuclear energy density optimization, Phys. Rev. C 82 (2010) 024313.
\newblock \href {https://doi.org/10.1103/PhysRevC.82.024313} {\path{doi:10.1103/PhysRevC.82.024313}}.

\bibitem{Kortelainen2012}
M.~Kortelainen, J.~McDonnell, W.~Nazarewicz, P.-G. Reinhard, J.~Sarich, N.~Schunck, M.~V. Stoitsov, S.~M. Wild, Nuclear energy density optimization: {L}arge deformations, Phys. Rev. C 85 (2012) 024304.
\newblock \href {https://doi.org/10.1103/PhysRevC.85.024304} {\path{doi:10.1103/PhysRevC.85.024304}}.

\bibitem{Kortelainen2014}
M.~Kortelainen, J.~McDonnell, W.~Nazarewicz, E.~Olsen, P.-G. Reinhard, J.~Sarich, N.~Schunck, S.~M. Wild, D.~Davesne, J.~Erler, A.~Pastore, Nuclear energy density optimization: {S}hell structure, Phys. Rev. C 89 (2014) 054314.
\newblock \href {https://doi.org/10.1103/PhysRevC.89.054314} {\path{doi:10.1103/PhysRevC.89.054314}}.

\bibitem{Goriely2005}
S.~Goriely, M.~Samyn, J.~M. Pearson, M.~Onsi, Further explorations of {S}kyrme--{H}artree--{F}ock--{B}ogoliubov mass formulas. {IV}. {N}eutron-matter constraint, Nucl. Phys. A 750 (2005) 425--443.
\newblock \href {https://doi.org/10.1016/j.nuclphysa.2005.01.009} {\path{doi:10.1016/j.nuclphysa.2005.01.009}}.

\bibitem{Papenbrock2024-pi}
T.~Papenbrock, \href{https://arxiv.org/abs/2410.00843}{Ab initio computations of atomic nuclei} (2024).
\newblock \href {http://arxiv.org/abs/2410.00843} {\path{arXiv:2410.00843}}, \href {https://doi.org/10.48550/arXiv.2410.00843} {\path{doi:10.48550/arXiv.2410.00843}}.
\newline\urlprefix\url{https://arxiv.org/abs/2410.00843}

\bibitem{Terashima2008-ky}
S.~Terashima, H.~Sakaguchi, H.~Takeda, T.~Ishikawa, M.~Itoh, T.~Kawabata, T.~Murakami, M.~Uchida, Y.~Yasuda, M.~Yosoi, J.~Zenihiro, H.~Yoshida, T.~Noro, T.~Ishida, S.~Asaji, T.~Yonemura, Proton elastic scattering from tin isotopes at {$295\,\mathrm{MeV}$} and systematic change of neutron density distributions, Phys. Rev. C 77 (2008) 024317.

\bibitem{Zenihiro2010-gk}
J.~Zenihiro, H.~Sakaguchi, T.~Murakami, M.~Yosoi, Y.~Yasuda, S.~Terashima, Y.~Iwao, H.~Takeda, M.~Itoh, H.~P. Yoshida, M.~Uchida, Neutron density distributions of {$^{204,206,208}\mathrm{Pb}$} deduced via proton elastic scattering at {$E_p=295\,\mathrm{MeV}$}, Phys. Rev. C 82 (2010) 044611.

\bibitem{Zenihiro2018-rt}
J.~Zenihiro, H.~Sakaguchi, S.~Terashima, T.~Uesaka, G.~Hagen, M.~Itoh, T.~Murakami, Y.~Nakatsugawa, T.~Ohnishi, H.~Sagawa, H.~Takeda, M.~Uchida, H.~P. Yoshida, S.~Yoshida, M.~Yosoi, \href{https://arxiv.org/abs/1810.11796}{Direct determination of the neutron skin thicknesses in {$^{40,48}\mathrm{Ca}$} from proton elastic scattering at {$E_p = 295\,\mathrm{MeV}$}} (2018).
\newblock \href {http://arxiv.org/abs/1810.11796} {\path{arXiv:1810.11796}}, \href {https://doi.org/10.48550/arXiv.1810.11796} {\path{doi:10.48550/arXiv.1810.11796}}.
\newline\urlprefix\url{https://arxiv.org/abs/1810.11796}

\bibitem{Liverts2021}
E.~Z. Liverts, \href{https://arxiv.org/abs/2110.01977}{Analytic scattering parameters at low energies} (2021).
\newblock \href {http://arxiv.org/abs/2110.01977} {\path{arXiv:2110.01977}}, \href {https://doi.org/10.48550/arXiv.2110.01977} {\path{doi:10.48550/arXiv.2110.01977}}.
\newline\urlprefix\url{https://arxiv.org/abs/2110.01977}

\end{thebibliography}






\end{document}